\documentclass[fleqn,usenatbib]{mnras}

\usepackage{newtxtext,newtxmath}

\usepackage[T1]{fontenc}

\DeclareRobustCommand{\VAN}[3]{#2}
\let\VANthebibliography\thebibliography
\def\thebibliography{\DeclareRobustCommand{\VAN}[3]{##3}\VANthebibliography}

\usepackage{siunitx}
\usepackage{graphicx}	
\usepackage{amsmath}	

\title[Disc polar alignment in multiple stellar systems]{Precession and polar alignment of accretion discs in triple (or multiple) stellar systems}

\author[Ceppi et al.]{
Simone Ceppi$^{1}$\thanks{E-mail: simone.ceppi@unimi.it},
Cristiano Longarini$^{1}$, 
Giuseppe Lodato$^1$,
Nicolás Cuello$^{3}$,
Stephen H. Lubow$^{4}$
\\
$^{1}$Dipartimento di Fisica, Università degli Studi di Milano, via Celoria 16, 20133 Milano, Italy \\
$^{3}$Univ. Grenoble Alpes, CNRS, IPAG, F-38000 Grenoble, France \\
$^{4}$Space Telescope Science Institute, 3700 San Martin Drive, Baltimore, MD 21218, USA.\\}

\date{Accepted XXX. Received YYY; in original form ZZZ}

\pubyear{2023}

\begin{document}
\label{firstpage}
\pagerange{\pageref{firstpage}--\pageref{lastpage}}
\maketitle

\begin{abstract}
We investigate the mechanism of polar alignment for accretion discs in hierarchical systems (HSs) with more than two stars. In eccentric binary systems, low mass discs that are sufficiently tilted to the binary orbit align in a polar configuration with respect to the binary plane by aligning their angular momentum to the binary eccentricity vector. In HSs, secular evolution of the orbital parameters makes the eccentricity vector of the system precess with time. This precession undermines the stability of the polar orbit for accretion discs hosted in HSs. We analytically show that the binary criteria for polar alignment derived in the literature are necessary but not sufficient conditions for polar alignment in HSs. Then, we derive an analytical criterion for polar alignment in HSs. In general, we find that discs orbiting the innermost level of a HS can go polar. Conversely, radially extended discs orbiting the outer levels of a HS cannot polarly align and evolve as orbiting around a circular binary.
We confirm our findings through detailed numerical simulations. Also, our results are compatible with the observed distribution of disc-orbit mutual inclination.
Finally, we compare the observed distribution of disc inclinations in the binary and in the HS populations. Binaries host mainly coplanar discs, while HSs show a wide range of disc inclinations. We suggest that the wider range of inclinations in HSs results from the secular oscillation of their orbital parameters (such as Kozai-Lidov oscillations), rather than from a different initial condition or evolution between HSs and binaries.
\end{abstract}




\section{Introduction}
Star formation takes place in clustered environments, where pre-stellar objects dynamically interact with each other, forming the seeds of multiple stellar systems \citep{clarkeclust,Reipurth+2014,Offner+2022}. Surveys of star forming environments \citep{mckeeoz}, as well as numerical simulations \citep{bate09,bate+10,Bate18}, show that stars form from a sequence of accretion episodes, for which the angular momentum is randomly oriented. In such a scenario, at least some accretion discs are expected to be misaligned to the stellar orbital plane. Recently, observations have confirmed these  theoretical predictions \citep{Czekala+19}. For example, the systems GG Tau \citep{ggtau1,ggtau2,ggtau3}, KH 15D \citep{kh15d1,lodfacc}, HD142527 \citep{Casassus+13_HD142527,Price+18_HD142527} and GW Ori \citep{gwori1,gwori2,small} exhibit high relative misalignments.

The dynamics of misaligned circumbinary discs dynamics has been actively studied during the last few decades, both theoretically \citep{warp1,warp2,warp3,Bate+00,nixon+12,nixon+13} and numerically \citep{facchlod,alydust1}. 
An inclined gas disc around a circular binary tends to precess around the binary angular momentum vector. If the disc sound crossing time is short enough (i.e. the inner part of the disc communicates its precession to the disc outer edge efficiently, \citealt{warp2}) the disc rigidly precesses with a frequency given by \cite{Bate+00} \citep[see also][]{lodfacc}. 
In addition, because of viscosity, the disc tends to align to the binary orbital plane, with an alignment timescale that is of the order of the viscous time \citep{alignment1,alignment2,Bate+00,Lubow&Ogilvie00} eventually reaching the coplanar equilibrium configuration. 
Also, misalignment in circumbinary discs affects dust dynamics \citep{longadust,alydust2}, creating substructures such as rings. 

When the binary system is eccentric an additional stable configuration is possible. \cite{Aly+15} and \cite{polar1} showed that a circumbinary disc around an eccentric binary tends to align its angular momentum vector to the eccentricity vector of the stellar system, leading to a polar configuration. This mechanism happens if the initial inclination is above a critical value, that is a decreasing function of the eccentricity. Polar alignment is likely to occur for high initial inclination, high eccentricity \citep{polar2}, cold and low mass discs \citep{polar3}.

Although we would expect to see several accretion discs in a polar configuration, only a few have  been observed up to now. One  is the disc orbiting HD98800 B \citep{hdp1}. Also, 99 Her is a debris disk in a polar configuration \citep{Kennedy+12} that likely evolved from a gaseous accretion disc \citep{Smallwood+20}. In addition, recently \cite{Kenworthy+22} showed that the light curve of an eclipse in V773 Tau multiple stellar system can be explained with the presence of a highly inclined disc. Interestingly, HD98800 B (as well as V773) is not a pure binary system (for which the polar alignment theory has been developed), but it resides inside a hierarchical quadruple stellar system. A recent study by \cite{Martin+22} finds that the 
high multiplicity of HD98800 does not prevent polar alignment of the circumbinary disc.

As in the case of HD98800, young multiple stellar systems often comprise more than two stars arranged in hierarchical configurations \citep{Duchene&Kraus13, Moe&DiStefano17}. A hierarchical configuration is made of nested binary orbits in which each semi-major axis is much larger than the semi-major axes of its subsystems. 
This semi-major axes hierarchy allows an analytic perturbative analysis of the orbital evolution that results in a periodic exchange of angular momentum between the orbits. As a result, hierarchical systems show periodic oscillations in the orbital shape and orientation \citep{Naoz+13}. Given this secular orbital evolution, the possibility of polar discs in systems with more than two stars needs to be further investigated.

In this work, we study whether discs in hierarchical systems can align polarly or if they fail to catch up with the secular evolution of the orbit. In section \ref{sec:analytics} we present the analytical framework under which we derive an additional criterion for the possibility of polar alignment of circumbinary discs in hierarchical systems. In Section~\ref{sec:numer} we numerically test the analytical findings of Section~\ref{sec:analytics}. In Section~\ref{sec:disc} we discuss the results of sections \ref{sec:analytics} and \ref{sec:numer}. We conclude in Section~\ref{sec:concl}.

\section{Polar configuration stability in systems with more than two stars}
\label{sec:analytics}

\begin{figure}
    \centering
	\includegraphics[width=0.451\textwidth]{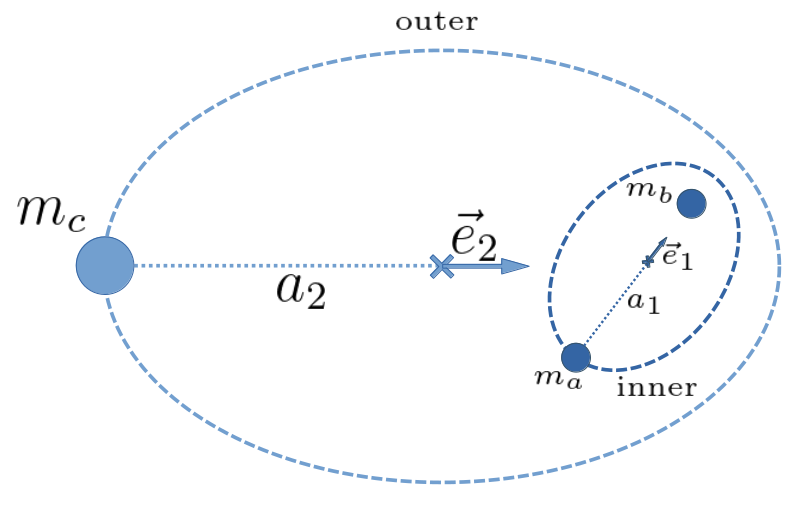}
    \caption{Hierarchical triple system sketch (not to scale). Light blue refers to the outer orbit, dark blue to the inner one. Dashed curves, crosses and arrows are the orbit, center of mass and eccentricity vector of the hierarchical level considered.}
    \label{fig:3sketch}
\end{figure}

\subsection{Disc polar alignment in systems with more than two stars}
We describe the orientation of a generic misaligned orbit around a central binary with the tilt angle, $\beta$, and the longitude of the ascending node (relative to the eccentricity vector direction), $\Omega$. The $\beta$ and $\Omega$ angles describe the direction of the specific orbital angular momentum vector (${\bf l}$) relative to the central binary angular momentum (${\bf l}_b$) and eccentricity vector (${\bf e}_b$). The frame of reference is defined by the instantaneous direction of the binary angular momentum and eccentricity vector with axes ${\bf e}_b$, ${\bf e}_b\times {\bf l}_b$ and ${\bf l}_b$. We define $\beta$ and $\Omega$ via the following relation:
\begin{equation}
    {\bf l} = (\sin{\Omega}\sin{\beta}, \cos{\Omega}\sin{\beta}, \cos{\beta}).
\end{equation}
By virtue of this definition, $\beta$ is the angle between the circumbinary orbit orbital plane and the binary orbital plane, while $\Omega$ is the angle between the direction of the binary eccentricity vector and the circumbinary orbit ascending node. Using these angles, we can describe the orientation of the orbital plane of a third body around a binary, as well as the orientation of a circumbinary accretion disc. In the latter case, $\beta$ and $\Omega$ can be functions of the disc radius $R$. In the following, we suppose the accretion disc to precess as rigid body, with a tilt and longitude of the ascending node independent of $R$.

Analytical studies of the restricted three body problem in \cite{Farago&Laskar10} showed that an inclined test particle orbiting an eccentric binary system undergoes two alternative kinds of precession. In the first kind, the angular momentum vector of the particle precesses around the angular momentum vector of the binary. 
For a circular orbit binary,  this kind of particle orbit,  known as a circulating orbit, maintains a constant angle between the particle orbital plane and the binary orbital plane (i.e. a constant tilt $\beta$), while the longitude of the ascending node $\Omega$ spans $2\pi$ with a typical frequency, given for example by \citet{lodfacc} (see their eq. 12). 
For an eccentric orbit binary, the tilt angle $\beta$ and nodal precession rate vary in time by an amount that depends on the binary eccentricity  \citep{Smallwood2019}. 
In the second kind of precession, the angular momentum of the particle precesses around the eccentricity vector of the binary. In this case, which is a librating orbit, $\beta$ oscillates around 90 degrees, while $\Omega$ oscillates around a fixed value. The orbit of the test particle belongs to the first or the second kind of orbit depending on the angles $\beta$ and $\Omega$, and on the eccentricity of the central binary $e_{\rm b}$. If the tilt of the particle is larger than a critical angle $\beta_{\rm crit}(e_{\rm b}, \Omega)$ the particle will precess around the binary eccentricity vector, otherwise around the binary angular momentum vector. For an orbit co-rotating with the central system, the critical angle for polar alignment is\footnote{for a counterrotating orbit $\beta_{\rm crit}(e_{\rm b}, \Omega) = \pi- \arcsin{\sqrt{\frac{1-e_{\rm b}^2}{1-5e_{\rm b}^2\cos{\Omega}^2+4e_{\rm b}^2}}}$} \citep[][]{Farago&Laskar10,Cuello&Giuppone19}
\begin{equation}
    \beta_{\rm crit}(e_{\rm b}, \Omega) = \arcsin{\sqrt{\frac{1-e_{\rm b}^2}{1-5e_{\rm b}^2\cos{\Omega}^2+4e_{\rm b}^2}}}.
    \label{eq:betacrit}
\end{equation}

A gaseous disc around an eccentric binary follows a similar evolution, except that, due to viscous dissipation, the orbital plane of the disc decays towards one of the two equilibrium configurations. Orbits precessing around the binary angular momentum tend to become coplanar (aligned with the binary orbital plane), while orbits precessing around the eccentricity vector decay towards a polar configuration (perpendicular to the orbital plane). Disc polar alignment in circumbinary discs has been studied in many previous works both analytically and numerically \citep[e.g.][]{Aly+15, polar1, polartimes, Cuello&Giuppone19, Zanazzi&Lai18}. 

Numerical simulations of the collapse of a molecular cloud show that the stellar and accretion disc formation process is very chaotic. In particular, we expect a randomly distributed initial misalignment between the stellar systems orbital planes and the discs orbiting them \citep{Bate18}. Also, from analytical investigations we expect no correlation between the disc and the binary initial angular momentum \citep{Toci+18}. Misaligned discs can result also from stellar flybys, which can generate an inclination up to 60$^\circ$ \citep[][]{Xiang-Gruess16, Cuello+2022}.
Thus, even if the conditions for polar alignment should be easily met, up to now we observed only one polarly aligned disc and it orbits around the inner binary of a hierarchical quadruple system \citep[i.e. HD~98800,][]{hdp1,Zuniga-Fernandez+21}. In addition, the vast majority of observed misaligned discs (particularly highly misaligned ones, that could undergo polar alignment) are found around systems with more than two stars \citep{Czekala+19}.

In general, stable stellar systems with more than two stars are arranged in nested binary orbits in which each semi-major axis is much larger than the semi-major axes of its subsystems. These configurations are called hierarchical and are the only ones that guarantee long-lasting stability to the stellar system. For example, fig.~\ref{fig:3sketch} sketches a triple system in a hierarchical configuration. Each hierarchical system level can be approximated to zeroth order by a binary system. Doing so, we could study polar alignment of discs around systems with more than two stars with the formalism developed by \citet{polartimes}. 
This crude approximation, however, completely neglects the perturbative term introduced in the gravitational potential by the presence of more than two stars. Taking into account this term means to reduce the hierarchical system to a hierarchical triple system. The exchange of angular momentum between the hierarchical levels of the system triggers further dynamical mechanisms such as Kozai-Lidov oscillations \citep{Zeipel10,Kozai62,Lidov62,Naoz16,Hamers21}.

In a hierarchical triple system, we can analytically compute the evolution of the binary orbital parameters of both the inner and outer orbit. Due to their mutual torques, we expect these osculating orbital elements to evolve with time on a secular timescale. In particular, the eccentricity vector direction, along which highly inclined discs would like to align their angular momentum, varies with time.

In coplanar hierarchical triple systems, we can analytically compute the precession frequency of the outer and inner orbit eccentricity vector. At the same time, a nearly polar aligned disc has a typical tilt oscillation frequency that is its nodal libration frequency \citep[][]{polartimes}. In the triple case the secular evolution of the orbital parameters shifts the polar orbit position with time, possibly undermining the stability of the polar configuration. In particular, if the nodal libration frequency is not high enough, the disc will not be able to track the evolving stable polar orbit, failing to remain polarly aligned.

In a hierarchical triple system we have two possible accretion discs orbiting a multiple stellar system: the first one orbiting the inner binary of the triple and the second one orbiting the outer orbit of the triple system. In the next sections we discuss the timescales involved in the two cases and the stability of the polar orbit around both the outer (Sec. \ref{outer-anal}) and the inner (Sec. \ref{inner-anal}) orbit.

\begin{figure*}
    \centering
	\includegraphics[width=0.952\textwidth]{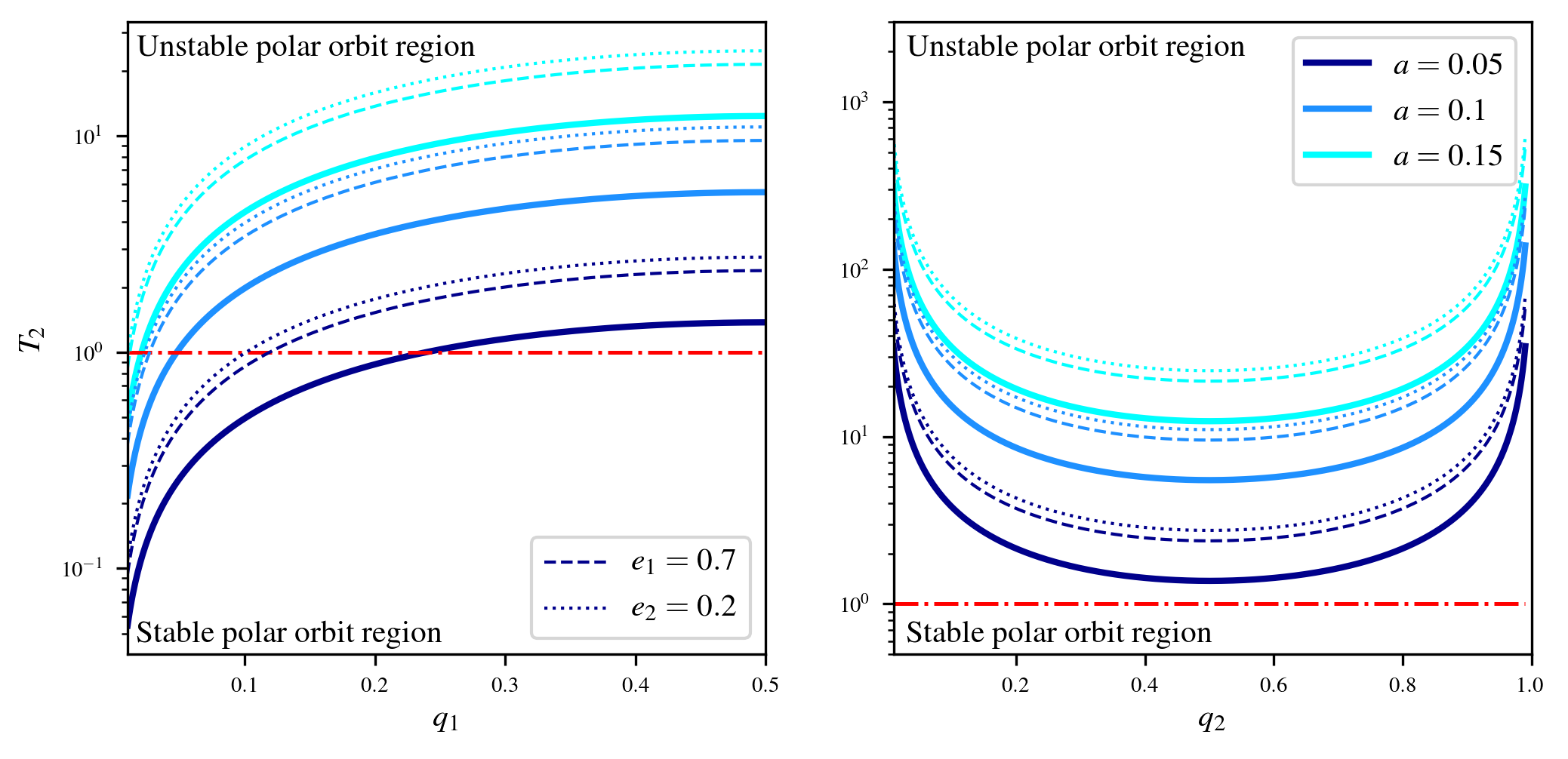}
    \caption{Dependency of $T_2$ (see sec. \ref{outer-anal}) on mass ratios, semi-major axes ratio, and eccentricities. When $T_2$ is smaller than unity (dash dotted red line) polar alignment is possible, conversely for $T_2>1$ it is not. On the left panel, $T_2$ is plotted as a function of the inner binary mass ratio $q_1$ (fixing $q_2$ to 0.5), while on the right one as a function of the outer binary mass ratio $q_2$ (fixing $q_1$ to 0.5). The thicker lines refer to $e_2=0.5$ and $e_1=0$. The dotted and dashed lines show how increasing $e_1$ to 0.7 (fixing $e_2=0.5$) and lowering $e_2$ to 0.2 (fixing $e_1=0$) affect $T_2$, respectively. Different colors refer to different semi-major axes ratios. }
    \label{fig:T2}
\end{figure*}

\subsection{Polar alignment in the circum-triple disc}
\label{outer-anal}

Let us consider a coplanar hierarchical triple as in the sketch of Fig. \ref{fig:3sketch}. The masses of the inner binary stars are $m_a$ and $m_b$, for the primary and the secondary respectively, while $m_c$ is the mass of the external third body. In addition, we define $q_1$, $e_1$, $a_1$ and $\Omega_1$ ($q_2$, $e_2$, $a_2$ and $\Omega_2$) the mass ratio, eccentricity, semi-major axis, and Keplerian frequency of the inner (outer) orbit. To zeroth order, the outer orbit of the triple is equivalent to the orbit of its associated binary, defined as the binary 
composed of $m_c$ and a single star of mass $m_a+m_b$ placed in the inner binary centre of mass. Thus, instantaneously, $m_c$ orbits the centre of mass of the inner binary along the same orbit as the one the associated binary would trace.

A nearly polar disc orbiting a pure binary undergoes nodal libration with the following theoretical frequency \citep{polartimes}:
\begin{equation}
    \omega_\text{pa} = \frac{3\sqrt{5}}{4}e_{\rm b}\sqrt{1+4e_{\rm b}^2}\frac{M_1 M_2}{(M_1+M_2)^2}\left<\left(\frac{a_{\rm b}}{R}\right)^{\frac{7}{2}}\right> \Omega_b,
    \label{eq:generalPA}
\end{equation}
where $e_b$ is the binary orbit eccentricity, $M_1$ and $M_2$ are the binary masses, $\Omega_b$ the binary Keplerian frequency and angular bracket notes the average of the ratio between the binary semi-major axis $a_b$ and the disc radius $R$, weighted over the angular momentum of the disc at radius $R$. 

We apply Eq. (\ref{eq:generalPA}) to the outer orbit case, as instantaneously it can be approximated by its associated binary. As for the masses, $M_1=m_c$ and $M_2=m_a+m_b$. For the eccentricity $e_b=e_2$ and for the orbital frequency $\Omega_b=\Omega_2$. We also compute the average weighted over the angular momentum assuming a Keplerian velocity field and that the disc extends from $R_{\rm in}$ to $R_{\rm out}$ with a density profile $\Sigma(R) \propto R^{-1}$. By doing so, we obtain:
\begin{equation}
    \omega_\text{pa,2} = \frac{9\sqrt{5}}{16}e_{2}\sqrt{1+4e_{2}^2}q_2(1-q_2)\left(\frac{a_2}{R_{\rm in}}\right)^{\frac{7}{2}}\frac{1-x_{\rm o}^{-2}}{x_{\rm o}^{3/2}-1} \Omega_2,
\end{equation}
where $x_{\rm o}=R_{\rm out}/R_{\rm in}$ and $q_2 = m_c/(m_a+m_b+m_c)$ is the outer orbit mass ratio.

The librating behaviour happens due to the precession of the disc angular momentum around the eccentricity vector of the triple outer orbit. However, the triple outer orbit eccentricity vector $\textbf{e}_2$ precesses with time due to the perturbation of the inner binary. The precession rate for the outer orbit can be analytically computed from the Hamiltonian of the stellar system with a perturbative approach \citep[e.g.][]{murrderm, Naoz+13}:
\begin{equation}
    \omega_{e_2} =\frac{3}{4} a^2 {q_1}(1-q_1) \frac{1+\frac{3}{2}e_1^2}{(1-e_2^2)^2} \Omega_2
    \label{out-precess-freq}
\end{equation}
where $a=a_1/a_2$ is the ratio between the inner and the outer orbit semi-major axis and $q_1 = m_b/(m_a+m_b)$ is the inner orbit mass ratio. This means that a polarly aligned disc around the outer orbit of a hierarchical triple system oscillates around a stable orbit that precesses with time with a frequency $\omega_{e_2}$. Thus, in order for the disc to stay ``attached" to the moving eccentricity vector, the condition for libration should be unaffected by the binary precession, meaning that $\omega_\text{pa,2}>\omega_{e_2}$.

To check the stability of small oscillations around the precessing polar configuration, we define the dimensionless quantity $T_2 = \omega_{e_2}/\omega_\text{pa,2}$: when $T_2<1$, polar alignment takes place; conversely, if $T_2>1$, it does not. The value of $T_2$ is
\begin{equation}
    \label{eq:T2}
    T_2 = \frac{4}{3\sqrt{5}}a^2\left(\frac{R_{\rm in}}{a_2}\right)^{\frac{7}{2}}\frac{x_o^{3/2}-1}{1-x_o^{-2}}\frac{q_1(1-q_1)}{q_2(1-q_2)}\mathcal{F}_2(e_1,e_2),
\end{equation}
where
\begin{equation}
    \mathcal{F}_2(e_1,e_2) = \frac{1+\frac{3}{2}e_1^2}{(1-e_2^2)^2e_2\sqrt{1+4e_2^2}}.
\end{equation}

The $T_2$ factor scales quadratically with the semi-major axes ratio: thus, either widening the outer orbit or shrinking the inner one decreases $T_2$ (which favours polar alignment). 
The torque between the inner and the outer orbit depends on $a$. Hence, the lower the value of $a$, the lower the precession frequency (i.e. the slower the precession).
In addition, $T_2$ depends on the inner and outer orbit eccentricities via the $\mathcal{F}_2$ term. This term diverges both for $e_2$ approaching unity and zero. In the former case, the precession frequency of the outer orbit diverges, making polar alignment impossible. In the latter case, the nodal libration frequency of a nearly polarly aligned disc reduces to zero. 
Indeed, as the outer binary becomes more circular, the process of polar alignment becomes less likely. 
A recent study by \cite{Lepp2023} examined the behavior of 
of circum-triple test particles and found a similar criteria as given by $T_2$.
Their criteria differs from $T_2$  mainly at small $e_2$.

As for the mass ratios, polar alignment is difficult for high $q_1$ and for $q_2$ approaching zero or unity. Indeed, higher values of $q_1$ (i.e. nearly equal mass inner binaries) translate into higher precession frequencies. Meanwhile, very high and very low $q_2$ reduce the outer binary to a single star concentrating the mass in the third body or in the inner binary (respectively) thus reducing the torque on the disc to zero. Conversely, when $q_1$ tends to zero, the inner binary perturbation becomes negligible --- slowing the precession. This favours polar alignment.

Figure \ref{fig:T2} shows how $T_2$ depends on the system orbital parameters. In this plot we assume $R_{\rm in}=1.5\ a_2$ (as expected in very inclined discs: \citealt{Lubow+15,Miranda&Lai15}) and $R_{\rm out}=30\ a_2$, thus $x_{\rm o}=30/1.5$. 
In the left panel, $T_2$ is plotted as a function of the inner binary mass ratio $q_1$ (fixing $q_2$ to 0.5, i.e. equal mass outer binary), while on the right one as a function of the outer binary mass ratio $q_2$ (fixing $q_1$ to 0.5, i.e. equal mass inner binary). The thicker lines refer to a configuration in which $e_2=0.5$. Such outer orbit eccentricity avoids extreme values that are expected to undermine (also) the stability of the triple stellar system and it sets the easiest configuration for polar alignment to occur, being $e_2=0.5$ the minimum of $\mathcal{F}_2$. Moreover, for $0.2<~e_2<~0.8$ the impact on $\mathcal{F}_2$ is of the order of 1. We also set $e_1=0$ to favour polar alignment as well. The dotted and dashed lines show how increasing $e_1$ to 0.7 (fixing $e_2$ to 0.5) and lowering $e_2$ to 0.2 (fixing $e_1$ to zero) affect $T_2$, respectively. The vast majority of the parameter space does not allow for polar alignment, as the curves are always above unity. The only exception is where the outer orbit is reduced to a two-body system, that is when $q_1$ tends to zero or for extremely small semi-major axis ratios $a$.

\begin{figure}
    \centering
	\includegraphics[width=0.451\textwidth]{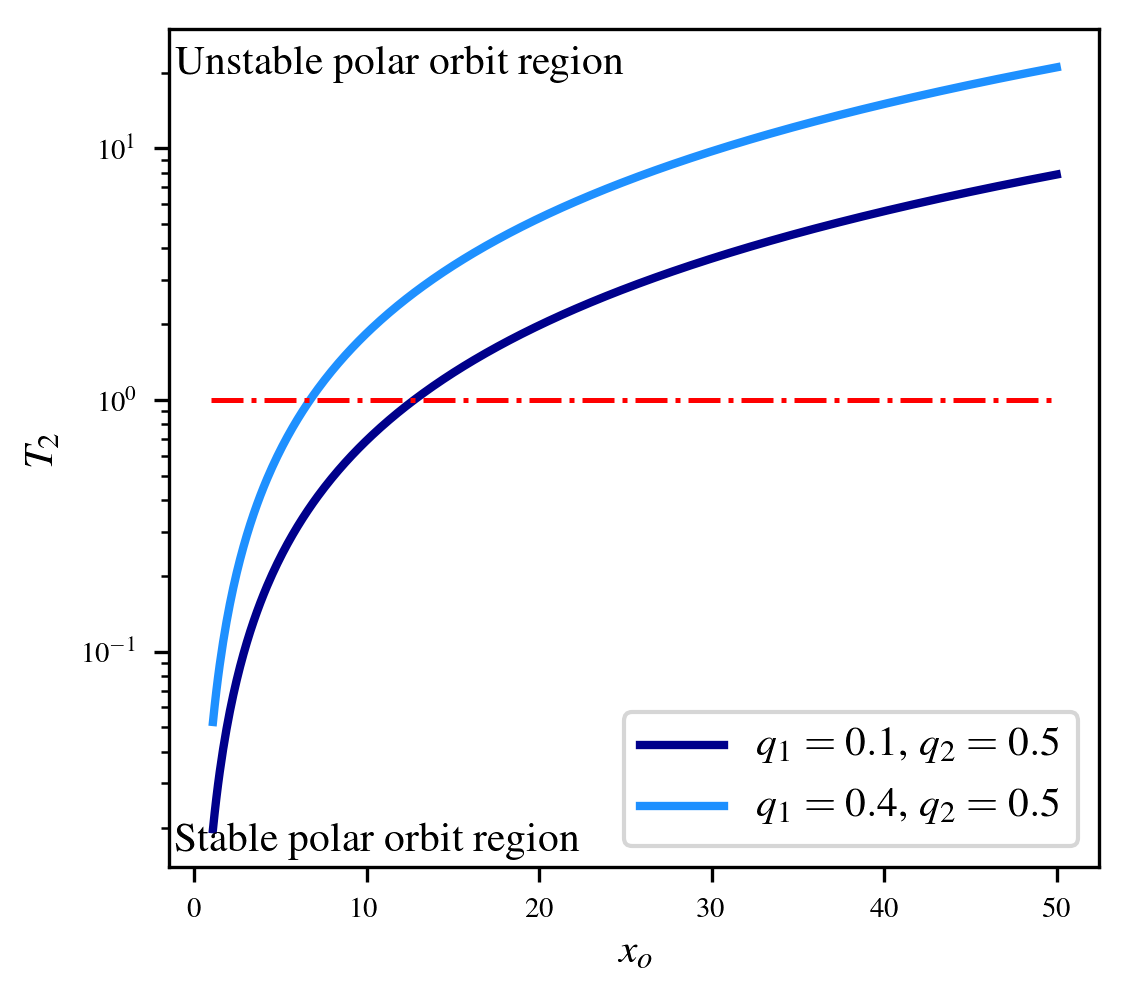}
    \caption{$T_2$ factor as defined in equation (\ref{eq:T2}) as a function of the ratio between the outer and the inner disc radii $x_o$. We used the same parameters of the solid line in figure \ref{fig:T2} with a semi-major axis ratio of 0.1. Different colors are different inner binary mass ratios. The outer binary mass ratio is fixed to 0.5, the most favourable to polar alignment.}
    \label{fig:T2xo}
\end{figure}

Last, the size of the disc also impacts the possibility of polar alignment. The more extended the disc the higher $T_2$, due to the lower nodal libration frequency of a radially extended disc. 
Figure \ref{fig:T2xo} shows the dependence of $T_2$ on $x_o$. The two curves refer to two different inner binary mass ratios (the outer binary mass ratio is set to $0.5$ the most favourable to polar alignment). Provided $x_o$ is low enough ($\lesssim 10$), the circum-triple accretion disc can become polar. Depending on the semi-major axis and mass ratios, if the triple is not isolated and the disc is truncated from the outside, $x_o$ can drop under that value, favouring the polar alignment of the disc.

\subsection{Polar alignment in the inner circum-binary disc}
\label{inner-anal}

The stability of the polar orbit for a disc orbiting the inner binary of a hierarchical triple system is determined by the inner binary eccetricity vector precession rate. In addition, \citet{Martin+22} showed that in this configuration also Kozai-Lidov oscillation of the disc (due to the interaction with the outer third body) could prevent circum-inner binary discs polar alignment. The disc Kozai-Lidov oscillation timescale can be computed by the theory of rigid disks \citep{coplanar2,Lubow&Ogilvie01} as done in \citet{Martin+14} (see their Eq. (4)). Using the same technique as in Equation~(\ref{eq:generalPA}), we obtain the following frequency for the circum-inner disc Kozai-Lidov oscillation:
\begin{equation}
    \omega_{\rm KL}=\frac{3}{4}\frac{q_2}{1-q_2}a^3\left(1-e_2^2\right)^{-3/2}\left(\frac{R_{\rm in}}{a_1}\right)^{3/2}\frac{1-x_{\rm o}^{-2}}{x_{\rm o}^{3/2}-1}\Omega_1 \,\,.
    \label{KLfreq}
\end{equation}

Firstly, we compare the disc Kozai-Lidov oscillation timescale with the other relevant ones. We compute the nodal libration frequency of a nearly polar orbit around the inner binary of a hierarchical triple system ($\omega_\text{pa,1}$) and the precession frequency of the inner binary eccentricity vector ($\omega_{e_1}$), as we did for the outer orbit case in Section~\ref{outer-anal}. We obtain:
\begin{equation}
    \omega_\text{pa,1} = \frac{9\sqrt{5}}{16}e_{1}\sqrt{1+4e_{1}^2}q_1(1-q_1)\left(\frac{a_1}{R_{\rm in}}\right)^{\frac{7}{2}}\frac{1-x_{\rm o}^{-2}}{x_{\rm o}^{3/2}-1} \Omega_1 \,\,,
    \label{omegapa}
\end{equation} 
and
\begin{equation}
    \omega_{e_1} = \frac{3}{4}a^3\frac{q_2}{1-q_2} \sqrt{\frac{(1-e_{1}^{2})}{(1-e_{2}^{2})^{3}}} \Omega_{1} \,\,.
    \label{omega1}
\end{equation}

\begin{figure}
    \centering
	\includegraphics[width=0.451\textwidth]{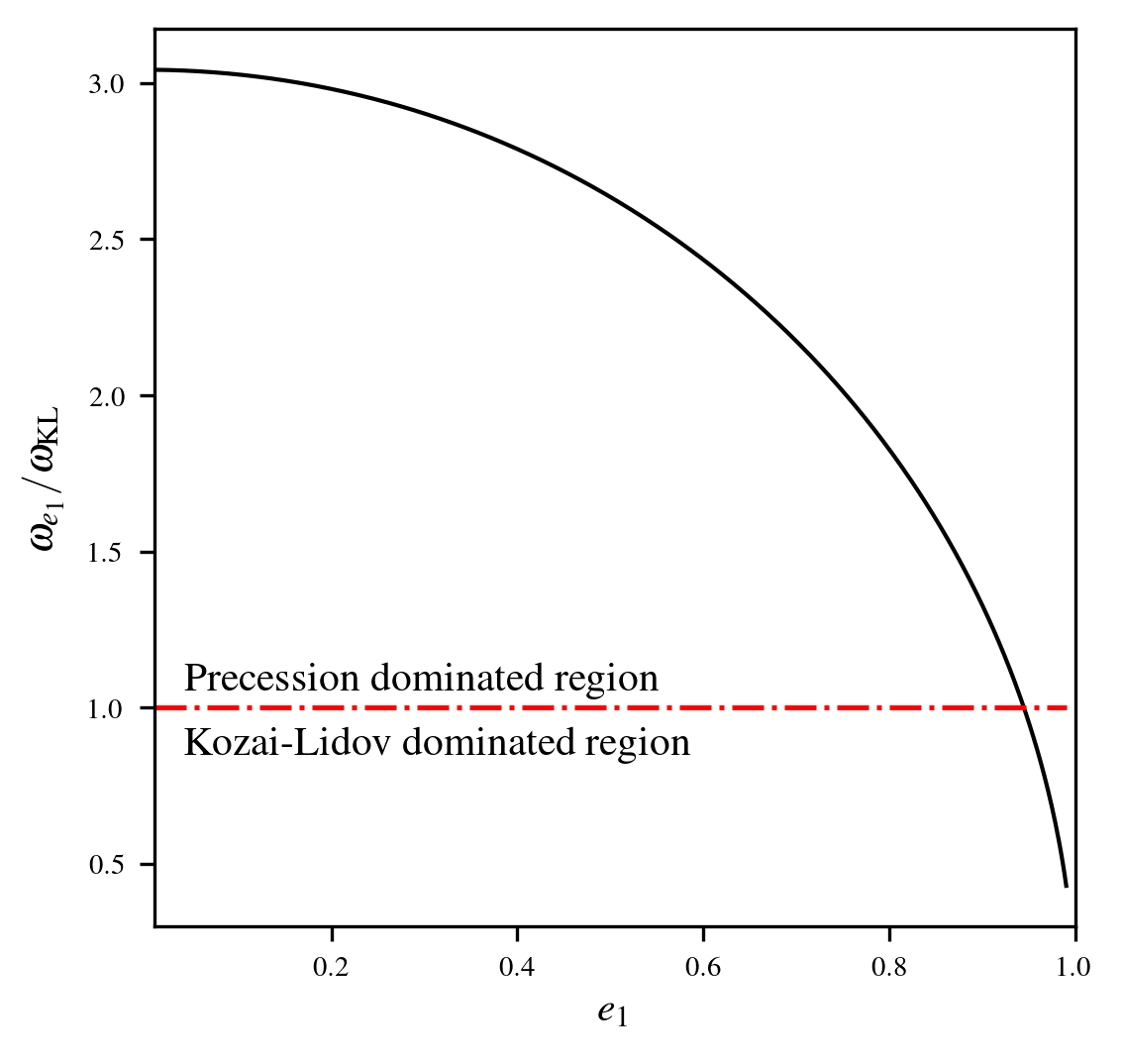}
    \caption{Ratio between the inner binary precession frequency and the disc Kozai-Lidov oscillation frequency. We use the same disc parameters as in Fig.~\ref{fig:T1} ($x_{\rm out} = 5/1.5$ and $R_{\rm in}/a_1=1.5$). The ratio is always above unity (red dash-dotted line) except for inner binary eccentricity near unity. Thus, in this configuration precession is always faster than disc Kozai-Lidov oscillation.}
    \label{fig:K1}
\end{figure}

Figure~\ref{fig:K1} shows the ratio between the precession frequency of the inner binary eccentricity vector ($\omega_{\rm e_1}$) and the disc Kozai-Lidov frequency of the circum-inner binary disc ($\omega_{\rm KL}$). This ratio solely depends on the eccentricity of the inner orbit $e_1$ and on the disc extent, as shown by Eqs.~(\ref{KLfreq}) and (\ref{omega1}). In Figure~\ref{fig:K1} we use $x_{\rm out} = 5/1.5$ and $R_{\rm in}/a_1=1.5$. With such parameters, apart from $e_1\sim1$, precession is always faster than disc Kozai-Lidov oscillation. In general, for $\omega_{\rm KL}>\omega_{\rm e_1}$ we would need a combination of extremely eccentric inner binaries, small discs ($x_{\rm out}<5/1.5$) and large disc cavities ($R_{\rm in}/a_1>2$). Thus, if polar alignment occurs because of a sufficiently slow precession of the inner binary, the disc Kozai-Lidov oscillation will not prevent polar alignment, since it is generally slower than precession. Having checked that, we go back to discussing the role of the inner binary precession. We define the corresponding $T_1 = \omega_{e_1}/\omega_{\text{pa},1}$ parameter for the inner binary as
\begin{equation}\label{eq:T1}
    T_1 = \frac{4}{3\sqrt{5}}a^3\left(\frac{R_{\rm in}}{a_1}\right)^{\frac{7}{2}}\frac{x_o^{3/2}-1}{1-x_o^{-2}}\frac{q_2/q_1}{(1-q_1)(1-q_2)}\mathcal{F}_1(e_1,e_2),
\end{equation}
with
\begin{equation}
    \mathcal{F}_1(e_1,e_2) =\sqrt{\frac{1-e_1^2}{(1-e_2^2)^3}} \frac{1}{e_1\sqrt{1+4e_1^2}},
\end{equation}

\begin{figure*}
    \centering 
	\includegraphics[width=0.952\textwidth]{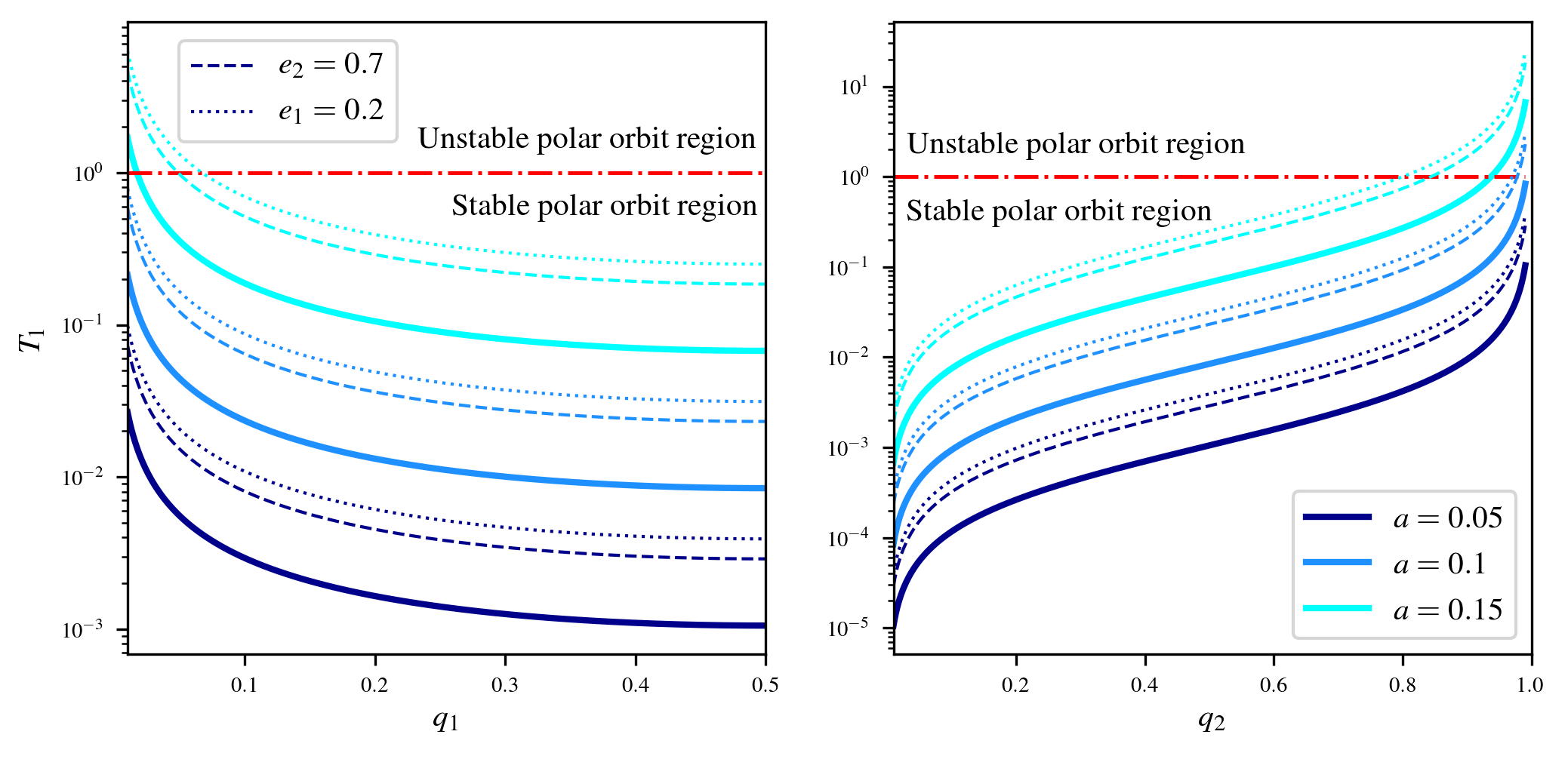}
    \caption{Dependency of $T_1$ (see sec. \ref{inner-anal}) on mass ratios,  semi-major axes ratio and eccentricities. When $T_1$ is smaller than unity (dash dotted red line) polar alignment is possible, conversely for $T_1>1$ it is not. On the left panel, $T_1$ is plotted as a function of the inner binary mass ratio $q_1$ (fixing $q_2$ to 0.5), while on the right one as a function of the outer binary mass ratio $q_2$ (fixing $q_1$ to 0.5). The thicker lines refer to $e_2=0$ and $e_1=0.5$. The dotted and dashed lines show how lowering $e_1$ to 0.2 (fixing $e_2=0$) and raising $e_2$ to 0.7 (fixing $e_1=0.5$) affect $T_1$, respectively. Different colors refer to different semi-major axes ratios.}
    \label{fig:T1}
\end{figure*}

The $T_1$ factor scales with the third power of the semi-major axes ratio. Thus, either widening the outer orbit or shrinking the inner one, results in an overall decrease of $T_1$. The precession frequency lowers (i.e. the precession is slower) with lower $a$ because the torque between the inner and the outer orbit weakens. In addition, $T_1$ depends on the inner and outer orbit eccentricities via the $\mathcal{F}_1$ term. This term diverges for both $e_2$ approaching unity and $e_1$ approaching zero. In the former case, the precession frequency of the inner orbit diverges, making polar alignment impossible. In the latter case, the nodal libration frequency of the polar disc tends to zero. 
Again, as the inner binary becomes more circular, the process of polar alignment becomes less likely.

As for the mass ratios, polar alignment is made difficult for high values of $q_2$ and for $q_1$ approaching zero. Higher values of $q_2$ result in a higher precession frequency. Meanwhile, very low $q_1$ reduce the inner binary to a single star, reducing the torque on the disc to zero. Conversely, when $q_2$ tends to zero, the third outer body perturbation becomes negligible, which slows the precession and favours polar alignment.

Figure \ref{fig:T1} shows how $T_1$ depends on the system orbital parameters. In this plot, we assume $R_{\rm in}=1.5\ a_1$ and $R_{\rm out}=5\ a_1$, thus $x_{\rm o}=5/1.5$. In the left panel, $T_1$ is plotted as a function of the inner binary mass ratio $q_1$ (fixing $q_2$ to 0.5), while on the right one as a function of the outer binary mass ratio $q_2$ (fixing $q_1$ to 0.5). The thicker lines refer to a configuration in which $e_1=0.5$ and $e_2 = 0$. The dotted and dashed lines show how raising $e_2$ to 0.7 (fixing $e_1$ to 0.5) and lowering $e_1$ to 0.2 (fixing $e_2$ to zero) affect $T_1$, respectively.

A disc orbiting the innermost binary of a hierarchical system is able to go polar in the vast majority of the inner binary parameter space, contrary to an accretion disc around the outer orbit of a hierarchical system. Except for very small $q_1$ and very high $q_2$, the $T_1$ factor is always below unity no matter the eccentricities, semi-major axis ratio or mass ratios.

In the case of HD98800 we have precise measurements of the orbital parameters of both the inner orbits and of the outer orbit of the quadruple stellar system \citep{Zuniga-Fernandez+21}. If we approximate HD98800 A to a single body, the disc around HD98800 B measures a $T_{1,{\rm HD}}\approx3\times10^{-4}$. This value is well below unity as expected, being the disc in a polar configuration.

\section{Numerical simulations}
\label{sec:numer}

We performed 3D numerical simulations using the Smoothed Particle Hydrodynamics (SPH) code {\sc Phantom} \citep{phantom}, widely used in the astrophysical community to study gas and dust in protostellar environments \citep[e.g.][]{Dipierro+2015,Mentiplay+2019,benniphantom,enriphantom,clauphantom,giuliaphantom,Vericel+2021} and for simulating the hydrodynamics of stellar systems with more than two stars embedded in accretion discs \citep[e.g.][]{Ragusa+2017,Ragusa+2021,polar1,Price+18_HD142527,Cuello+2019,Poblete+2019,Calcino+2019,small,Ceppi+22}.
These simulations are designed to test the criteria showed in Sections~\ref{outer-anal} and \ref{inner-anal} for polar alignment in discs orbiting a hierarchical triple outer (sec. \ref{outer-num}) and inner (sec. \ref{inner-num}) orbit. 

\subsection{Circum-triple disc simulations}
\label{outer-num}
Simulations \textbf{S1} and \textbf{S2} test the polar alignment of a disc around the outer orbit of a hierarchical triple (\textbf{S1}), compared to the same disc orbiting around the binary associated to the triple outer orbit (i.e. the binary with the same orbital parameters of the triple outer orbit and made of the third body and a star obtained by condensing the triple inner binary in its centre of mass, simulation \textbf{S2}). The triple system consists of a circular equal mass inner binary ($m_a=m_b=0.5~ M_\odot$) orbited by a coplanar third star with mass equal to the inner binary mass ($m_c=m_a+m_b=1~ M_\odot$). The semi-major axis of the inner binary is $a_1=1$~au. The semi-major axis and the eccentricity of the outer orbit are $a_2=10$~au and $e_2=0.5$. The outer and the inner orbital plane are coplanar. The associated binary system consists of a star of mass $m_c$ orbited by a star of mass $m_a+m_b$. The eccentricity and semi-major axis of the binary are the same as the triple outer binary orbit. 

Both the binary and triple system are orbited by the same gaseous accretion disc. The disc is initially tilted to the stellar orbital plane of $\beta=70^\circ$ and the longitude of the ascending node $\Omega$ equals $90^\circ$. Each simulated disc is made of 
$3\times10^5$
SPH particles, resulting in a vertical resolution of $\left<h/H\right>\sim0.24$. The total gas mass is initially the $1\%$ of the stellar mass and the discs extend from $R_{\rm in}=15$~au to $R_{\rm out}=100$~au. The gas surface density is initially distributed with a power law profile
\begin{equation}
    \Sigma(R)=\Sigma_{\rm in}\left(\frac{R}{R_{\rm in}}\right)^{-p}\left(1-\sqrt{\frac{R_{\rm in}}{R}}\right),
    \label{eq:surfdens}
\end{equation}
with $p=1$ and $\Sigma_{\rm in}=22.2$ g/cm$^2$. We adopt a locally isothermal equation of state $P=c_s^2\rho$, with
\begin{equation}
    c_{\rm s}(R)=c_{\rm s}(R_{\rm in})\left(\frac{R}{R_{\rm in}}\right)^{-q}, 
    \label{eq:temp}
\end{equation}
with $q=0.25$. The resulting disc aspect ratio is given by
\begin{equation}
    \frac{H}{R}=\frac{H_0}{R_0}\left(\frac{R}{R_{\rm in}}\right)^{1/4},
    \label{eq:aspect}
\end{equation}
with an $H_0/R_0=0.1$ at $R=R_{\rm in}$.
Viscosity is implemented with the artificial viscosity method \citep{Lucy77, Gingold&Monaghan77} resulting in a \citet{Shakura&Sunyaev73} $\alpha$-viscosity \citep{Lodato&Price10} with $\alpha_{\rm SS}\sim0.01$ ($\alpha_{\rm AV}=0.42$). In every simulation, stars are modelled as sink particles \citep{Bate+95, phantom}.

The initial tilt of the disc is well above the critical angle for polar alignment around a binary system, which is $\beta_{\rm crit}(e=0.5, \Omega=90^\circ)\sim40^\circ$ \citep{Farago&Laskar10}. Thus, we expect the circumbinary disc in simulation \textbf{S2} to polarly align. Conversely, even if the outer orbit parameters in the hierarchical triple of simulation \textbf{S1} are the same as the binary in \textbf{S2}, $T_2\sim6> 1$. Thus, we expect the disc not to align polarly.

\begin{figure}
    \centering
	\includegraphics[width=0.476\textwidth]{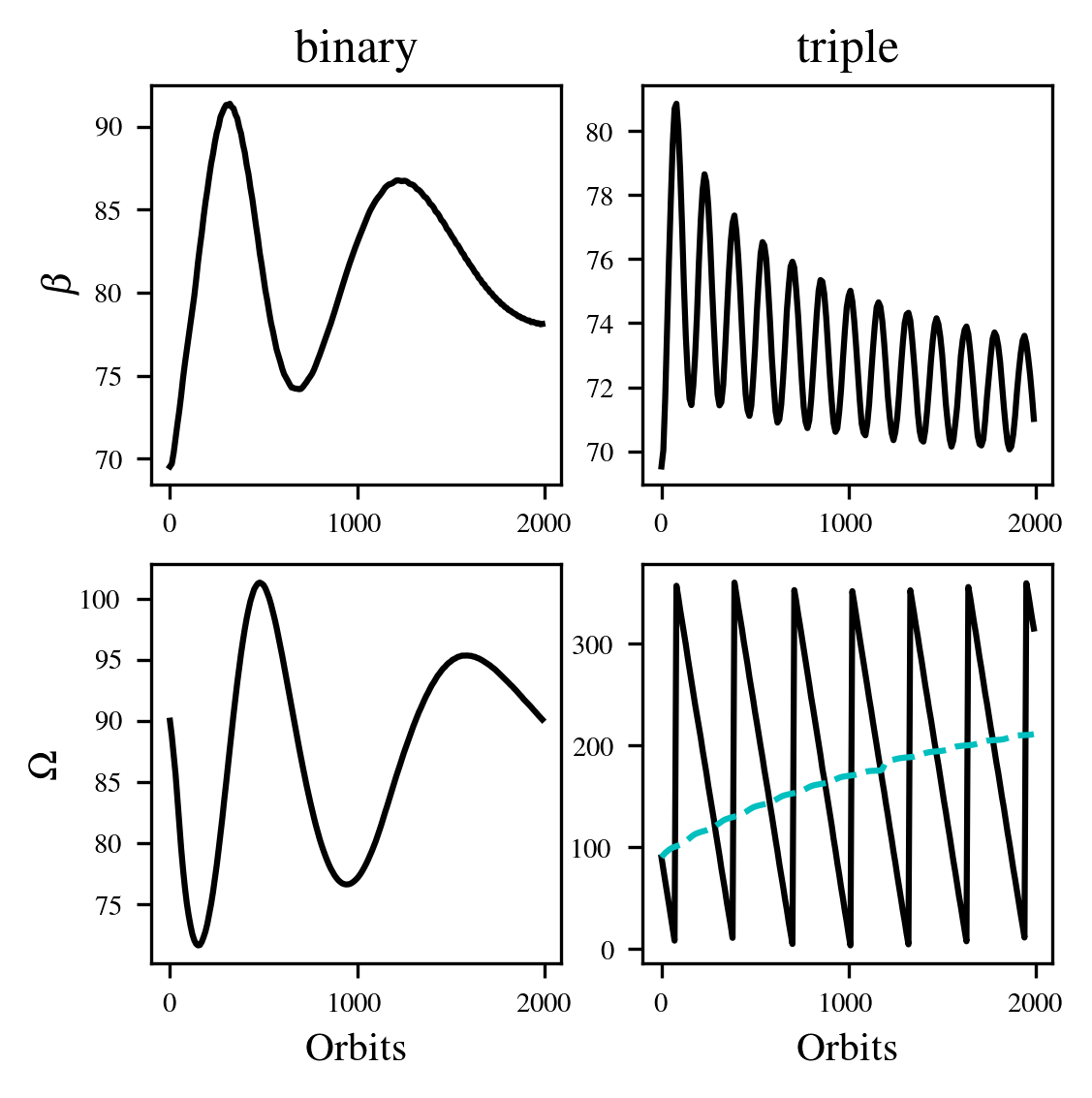}
    \caption{Radially averaged tilt (top row) and longitude of the ascending node (bottom row) of the circumbinary (left column, simulation \textbf{S2}) and circumtriple (right column, simulation \textbf{S1}) disc as a function of time. Dashed line is the longitude of the ascending node with respect to the initial semi-major axis direction, rather than the eccentricity vector. In the left column the circum-binary disc is aligning polarly. As expected due to the mass of the disc, the polar alignment inclination is less than 90 degrees.} In the right column the circum-triple disc is going coplanar, even if the orbital parameters of the triple outer orbit are the same as the binary orbital parameters.
    \label{fig:outer-num-comp}
\end{figure}

Figure \ref{fig:outer-num-comp} shows radially averaged tilt ($\beta$, top panels) and longitude of the ascending node ($\Omega$, bottom panels) profiles of the disc as a function of time. 
The two systems behave very differently: the circumbinary disc undergoes polar alignment, since the inclination oscillates around $\beta\sim 90^\circ$ and the longitude of the ascending node librates, as expected for a disc going polar \citep{polar1}. On the other hand, the circumtriple disc aligns to the stellar orbital plane (i.e. the inclination decreases) and it precesses, with its longitude of the ascending node spanning $2\pi$. There are two processes that make the longitude of the ascending node to precess: i) the eccentricity vector, to which the longitude of the ascending node refers, is precessing; ii) the disc itself is precessing, as an inclined circumbinary disc would around a pure binary. This can be seen computing the longitude of the ascending node referring to the initial eccentricity vector position. This absolute longitude of the ascending node is plotted as a dashed curve in Fig. \ref{fig:outer-num-comp}. As expected, polar alignment does not occur in this case, since the precession rate of the eccentricity vector is larger than the polar alignment one.

The disc inclination, apart from the oscillating behaviour, is exponentially decaying towards the mid-plane. We fit the exponential decay to find the timescale of coplanar alignment, obtaining $\tau_{\rm cop}\sim3.8\times10^4$ outer orbit periods. The timescale $\tau_{\rm cop}$ is of the order of the disc viscous timescale for this disc ($\sim2\times10^4$ outer orbits).
\cite{Bate+00} found that the alignment timescale (given by their equation (35), see also \cite{Lubow&Ogilvie00}) for a tilted disc is of the order of (or slightly longer than) the disc viscous timescale. This suggests that an accretion disc with $T_2>1$ orbiting the outer levels of a hierarchical system sees the central system as a circular binary.
In fact, the evolution of the tilt angle of the circum-triple disc in simulation \textbf{S1} is similar to a circum-binary disc orbiting a low eccentric binary system (an exponentially decaying tilt and a precessing longitude of the ascending node). This is due to the fact that the disc sees the eccentricity vector of the central stellar system averaged over the precession period. 

\subsection{Circum-binary disc simulation in a triple system}
\label{inner-num}
Simulation \textbf{S3} studies a disc orbiting the inner binary of a hierarchical triple system. The stellar system consists of an eccentric ($e_1=0.5$) equal mass inner binary ($m_a=m_b=0.5\,M_\odot$) orbited by a third star with mass equal to the inner binary mass ($m_c=m_a+m_b=1\,M_\odot$) in a circular orbit. The semi-major axis of the inner and outer binary are $a_1=1$~au and $a_2=20$~au respectively, and the inner and outer orbits are coplanar.

The inner binary is orbited by a circum-binary disc with an initial tilt of $\beta=70^\circ$ and a longitude of the ascending node of $90^\circ$. The disc is simulated with 
$3\times10^5$ SPH particles, resulting in a vertical resolution of $\left<h/H\right>\sim0.23$. The total gas mass is initially the 1\textperthousand \ of the inner binary mass and the disc extends from $R_{\rm in}=1.5$~au to $R_{\rm out}=5$~au. The gas surface density profile is the same as in Eq. (\ref{eq:surfdens}), with $\Sigma_{\rm in}=2.7\times10^2$ g\,cm$^{-2}$, and the temperature and aspect ratio profiles are given by Eqs. \ref{eq:temp} and \ref{eq:aspect}. Viscosity and stars are implemented as in previous simulations, with $\alpha_{\rm SS}\sim0.01$ ($\alpha_{\rm AV}=0.44$).

The initial disc tilt is above the critical angle for polar alignment around a pure binary system. In addition, in this configuration $T_1\sim0.008<1$, which should translate into disc polar alignment (see Sect.~\ref{inner-anal}). 

\begin{figure}
    \centering  
	\includegraphics[width=0.3\textwidth]{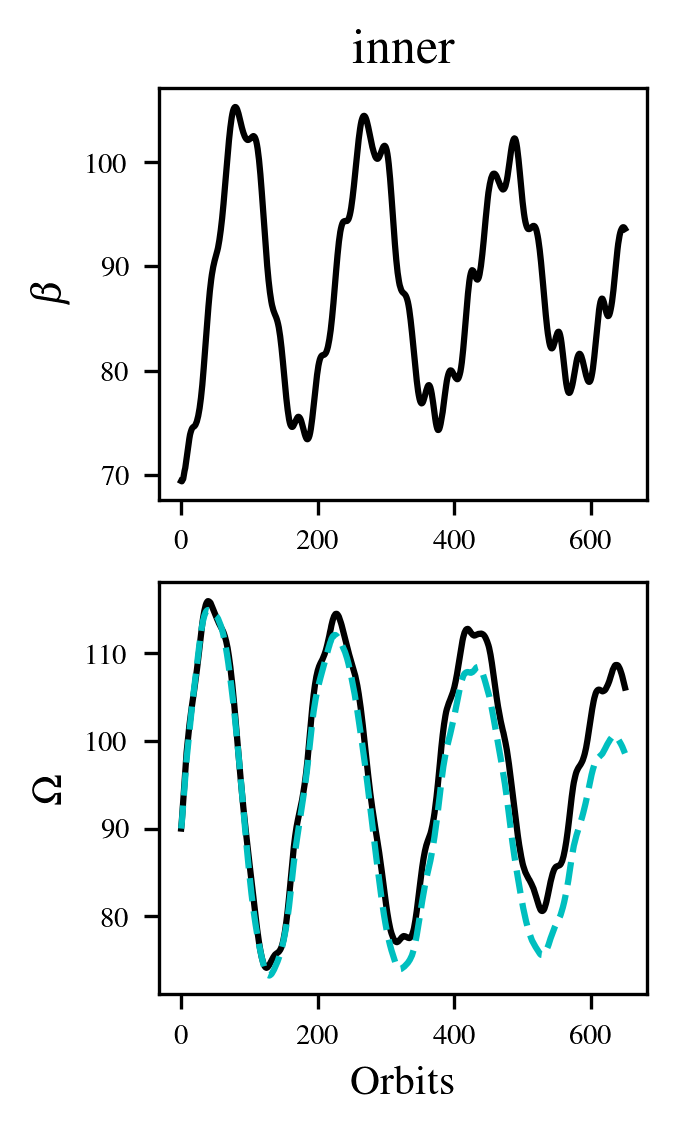}
    \caption{Radially averaged tilt (top row) and longitude of the ascending node (bottom row) of the circum-inner binary disc (simulation \textbf{S3}) as a function of time. Dashed line is the longitude of the ascending node with respect to the initial semi-major axis direction, rather than the eccentricity vector. The disc is aligning to a polar configuration. The libration timescale and the stationary inclination differ from the circumtriple disc due to the lower mass of the circuminner binary disc.}
    \label{fig:inner-num}
\end{figure}

Figure \ref{fig:inner-num} shows radially averaged tilt ($\beta$, top panel) and longitude of the ascending node ($\Omega$, bottom panel) profiles of the disc as a function of time. 
The tilt plot clearly shows that the disc oscillates around $90^\circ$ with the amplitude being damped over time. The longitude of the ascending node is librating as well: thus, the disc is going polar as expected. However, the disc is librating around the eccentricity vector of the inner binary, which in turn is precessing with time. As a consequence, the disc angular momentum is precessing as well. This can be seen computing the disc longitude of the ascending node with
respect to the initial semi-major axis direction, instead of referring to the inner binary eccentricity vector, which is precessing. This absolute longitude of the ascending node is plotted in the second row of Fig. \ref{fig:inner-num} with a dashed curve. The absolute longitude of the ascending node is oscillating as the relative longitude of the ascending node, but its mean value is decreasing with time following the precession of the inner binary eccentricity vector.

\section{Discussion}
\label{sec:disc}

\subsection{Polar alignment in hierarchical stellar systems}
\label{sec:disc1}
Hierarchical systems with more than two stars are common in a young stellar population \citep{Duchene&Kraus13, Moe&DiStefano17}, and they typically host discs orbiting their hierarchical levels. A significant fraction of misaligned stellar discs studied in the literature are orbiting or are inside systems with more than two stars \citep[e.g.][]{Phuong+20, Keppler+20,gwori1, gwori2,Price+18_HD142527,hdp1}. 
By studying the observed misalignment distribution of discs in multiple stellar systems, we aim at obtaining information about their physical properties, their formation and early stage evolution. Properties of discs (e.g. size, viscosity) affect the evolution of the tilt angle which, in return, provides information about them. In order to extract information from the final observed tilt distribution, we need to understand the way in which it evolves and, possibly, the initial conditions of its evolution. 

In particular, the fraction of discs in a polar configuration gives us insights about the fraction of highly-eccentric binary systems with highly-inclined discs. Remarkably, the time evolution towards a polar orbit is faster than the disc lifetime. So, in principle, the final outcome of the evolution process (i.e. the observed fraction of polar discs) directly traces the initial conditions (i.e. the outcome of star formation process). Indeed, the fraction of polarly aligned discs is related to how common the critical configuration is (i.e. a binary with eccentricity $e_{\rm b}$, orbited by a disc with tilt $\beta>\beta_{\rm crit}(e_{\rm b},\Omega)$). 
However, we have to take into account that pure binary systems and systems with more than two stars have different critical conditions for polar alignment.

The analytical and numerical findings, reported in Sections~\ref{sec:analytics} and \ref{sec:numer}, show that the secular evolution of the osculating orbital elements in hierarchical systems is a threat to disc polar alignment. In particular, a fast enough precession of the orbit eccentricity vector prevents the mechanism of polar alignment. Figures \ref{fig:T2} and \ref{fig:T1} show the $T$ parameter discussed in Sections \ref{outer-anal} and \ref{inner-anal} (respectively) as a function of the relevant parameters of a hierarchical triple (the two mass ratios $q_1$ and $q_2$, the semi-major axis ratio $a$ and the inner and outer orbit eccentricities $e_1$ and $e_2$). When $T$ is larger than unity, the polar orbit is unstable --- no matter how high the disc-orbit misalignment is. Under this condition, the pure binary criteria for polar alignment derived in the literature can not be applied to hierarchical systems.

Looking at the parameter space explored in Figure \ref{fig:T2} and at the numerical results in section \ref{outer-num}, discs orbiting the outer levels of hierarchical systems are generally not able to go polar. Indeed, the $T$ parameter is always larger than unity, except for radially narrow discs and where the system is reduced to an actual binary system (i.e. for a mass ratio of the inner binary ($q_1$) approaching zero and for very small semi-major axis ratios ($a$)). Thus, regardless of the mutual inclination, a misaligned radially extended disc orbiting an outer level of a hierarchical system always evolves towards a coplanar configuration. Since the eccentricity vector is quickly precessing, the disc evolves as orbiting a circular orbit given that the eccentricity vector is averaged over a precession period.

In a circum-binary disc in an hierarchical system, however, polar alignment is possible as the $T$ factor is generally below unity (see Figs. \ref{fig:T1} and \ref{fig:inner-num}). As the precession of the inner binary eccentricity vector is slow enough for the polar disc to follow it, the disc remains locked to the eccentricity vector, precessing with it and conserving the polar configuration. Note that the disc precession happens on the eccentricity vector precession timescale and not on the typical nodal precession timescale for a tilted circum-binary disc \citep{Bate+00,lodfacc}. Superimposed to this precession, the longitude of the ascending node is librating as well, due to the precession of the disc angular momentum vector around the inner binary eccentricity vector.

The evolution of the osculating elements in hierarchical systems further affects the polar alignment process in such systems. The critical angle for polar alignment significantly depends on $\Omega$, that is relative to the central orbit eccentricity vector (see eq. \ref{eq:betacrit}). Given that this vector in hierarchical system precesses with time, $\Omega$ is constantly spanned with time. Thus, regardless of the initial $\Omega$ value, the triple configuration will precess, eventually exploring the $\Omega$ related to the lowest critical angle. Therefore, even if the initial $\Omega$ results in a configuration where $\beta<\beta_{\rm crit}$ (so no polar alignment for a pure binary configuration), the triple configuration will polarly align as soon as the $\Omega$ precession allows it to fullfill the condition $\beta>\beta_{\rm crit}(\Omega, e_{\rm b})$. Thus, given the eccentricity of the orbit, the critical angle for polar alignment in hierarchical systems is always the one with $\Omega=90^\circ$ (the lower one).

Up to now, we only discussed hierarchical stellar systems with inner and outer orbits coplanar to each other. When the inner and the outer orbit of a hierarchical triple are misaligned, the secular evolution of the osculating orbital parameters is even more complex. Indeed, not only the argumentum of periapsis precesses with time, but also Kozai-Lidov oscillations of the stellar orbits
occur. Thus, the eccentricity of each hierarchical level orbit and the mutual inclination between the hierarchical orbital planes oscillate with time. However, at least for a circular outer orbit, the Kozai-Lidov oscillation frequency is similar to the precession frequency of the inner binary orbit we derived in Eq. (\ref{omega1}) \citep{Naoz16, Antognini15}.
Therefore, the $T$ parameter is still reliable to predict the possibility of polar alignment in the inner orbit of misaligned hierarchical triples. However, a thorough analysis of this problem is needed, especially regarding the possibility of polar alignment in misaligned hierarchical triples circum-outer orbit accretion discs and we defer this to subsequent work. An aspect of crucial importance is that Kozai-Lidov oscillations can trigger polar alignment in misaligned hierarchical systems. Indeed, in the oscillation the orbits eccentricities raise (lowering the critical angle) and the orbital inclinations oscillation makes also the tilt of the disc to change. This can trigger polar alignment in configurations with a lower-than-unity $T$ parameter, but a low initial orbit-disc misalignment.

We thus expect polar alignment to still be possible only around the innermost level of hierarchical systems even for misaligned multiple systems, where it can be even promoted by Kozai-Lidov oscillations and the $\Omega$ precession.

\subsection{Observed tilt distribution of discs around binaries and hierarchical systems}

\begin{figure}
    \centering
	\includegraphics[width=0.476\textwidth]{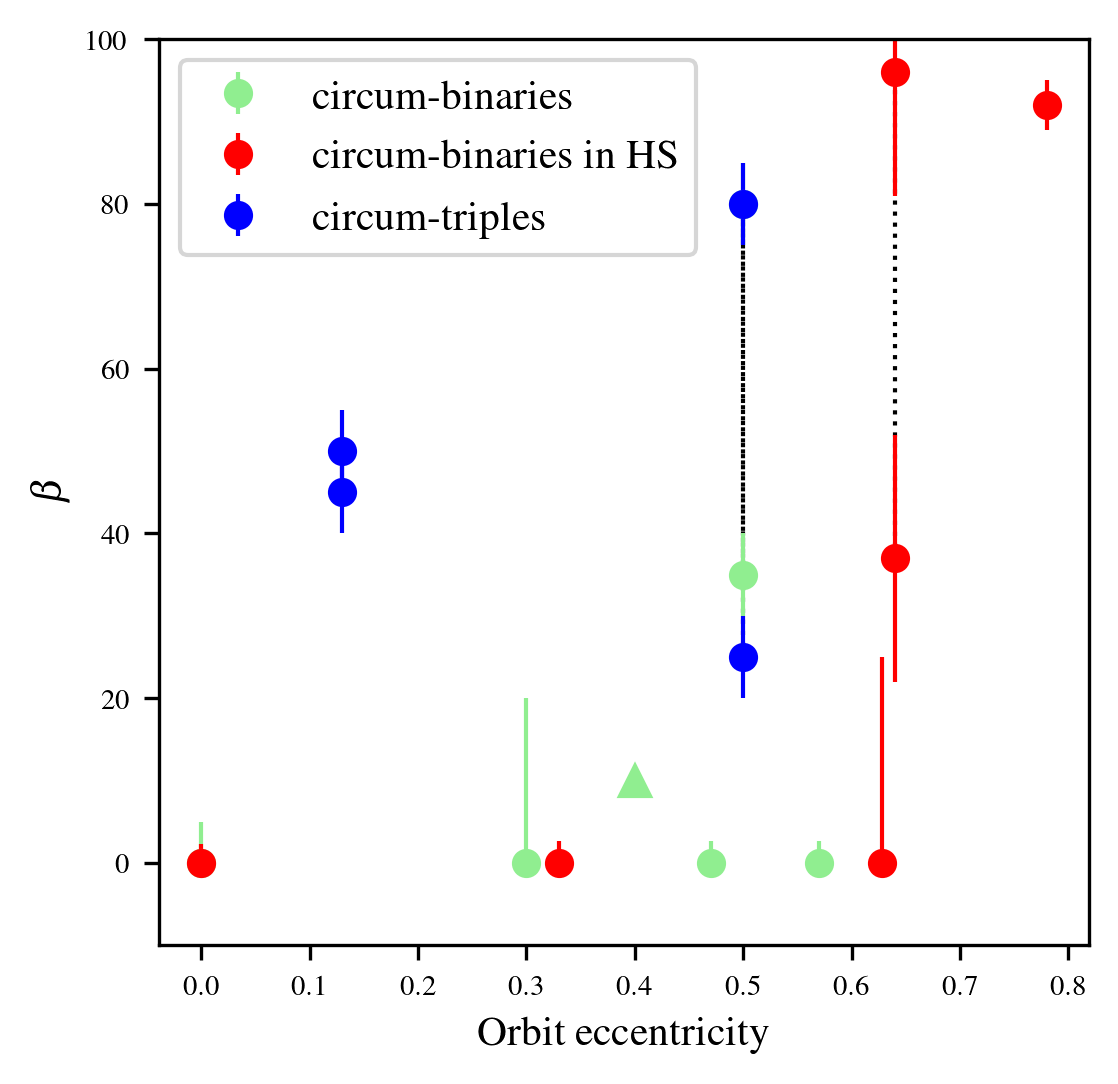}
    \caption{Disc-orbit misalignment of circumbinary discs in Tables 3 and 4 in \citet{Czekala+19} as a function of the central system eccentricity. Green dots are accretion discs orbiting pure binary systems, blue dots are accretion discs orbiting an outer hierarchical level of a hierarchical system, red dots are accretion discs orbiting the innermost hierarchical level of a hierarchical system (HS). The triangle represents the lower limit on the mutual inclination of R~Cra. Dotted lines connect the two degenerate solutions for HD~142527, SR~24N, and GG Tau A systems.}
    \label{fig:CzekalaPlot}
\end{figure}

\begin{figure}
    \centering
	\includegraphics[width=0.476\textwidth]{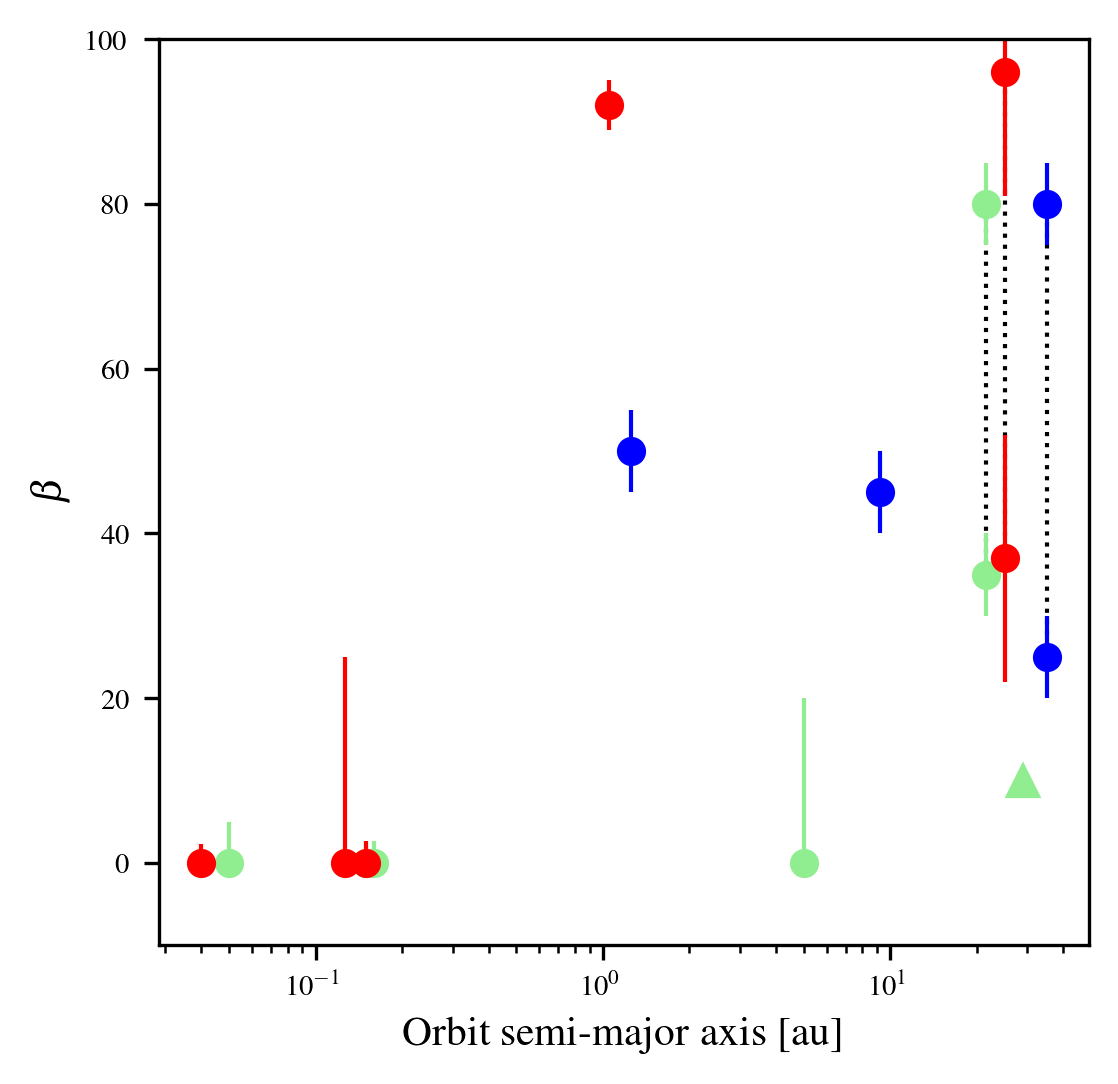}
    \caption{Disc-orbit misalignment of circumbinary discs in Tables 3 and 4 in \citet{Czekala+19} as a function of the orbit semi-major axis. Legend is the same as figure \ref{fig:CzekalaPlot}.}
    \label{fig:CzekalaPlotsma}
\end{figure}

Figure~\ref{fig:CzekalaPlot}, originally reported by \cite{Czekala+19}, collects all protoplanetary discs in the literature orbiting in multiple stellar systems for which we know the inclination between the disc and the system orbital plane. Each point in the plot represents a disc orbiting a pair of bound stars, either in a pure binary system or in a higher multiplicity system. Green dots refer to discs orbiting in a pure binary system, blue dots refer to discs that have more than two stars orbiting in their cavity (i.e. the outer level of hierarchical systems) and red dots refer to discs orbiting a binary with external companions outside the disc (i.e. the innermost level of hierarchical systems). The dots connected by dotted vertical lines are two possible solutions for the mutual inclination due to a 180° ambiguity in the longitude of the ascending node of the system due to lack of radial velocity information. 

The three populations show different tilt distribution. Discs orbiting pure binary systems (green dots) are mostly coplanar, hence there are no polarly aligned discs. This mismatches with the theoretical expectation. If the initial tilt distribution is nearly randomly distributed, we expect a wider scatter in the distribution because the population should be slowly going coplanar on a tilt evolution timescale comparable to the viscous timescale \citep{Bate+00, Lubow&Ogilvie00}. This could be due to a lower average tilt in the initial population or to a faster than expected tilt evolution (e.g. due to an higher than expected viscosity). In addition, we lack the expected small population of polarly aligned discs (resulting from the fraction of discs initially more tilted than the critical angle for polar alignment in binaries). Hence, either the conditions for polar alignment are less populated than expected or external factors reduce the stability of the polar configuration (e.g. the interaction with the environment).

The tilt distribution of circum-binary discs in hierarchical systems is more articulated. We have a small fraction of polarly aligned discs (i.e. the disc in HD98800B and possibly in SR~24N), although there are less discs in this populations than in the pure binaries one. This is in agreement with our analytical findings (polar configuration is stable in such systems) and suggests that the conditions for polar alignment are more likely for those systems than in pure binaries. This is also in line with what discussed in section \ref{sec:disc1}: phenomena that are typical of systems with more than two stars could be able to foster polar alignment. First, polar alignment for  such configurations could be triggered by Kozai-Lidov oscillations of the hierarchical system. Second, by varying the longitude of the ascending node, the precession of the orbit eccentricity vector allows the disc to always reach the configuration in which the critical angle for polar alignment is minimum. We defer the study of the impact of these mechanisms on polar alignment to future works. Additionally, depending on the stellar system parameters (such as binary mass, semi-major axis, eccentricity) general relativity could limit the possibility of polar alignment for extended discs \citep{Lepp+22}. Thus, the presence of an additional external companion truncating the disc from the outside could facilitate polar alignment for circum-inner binary discs as well. The rest of the population is nearly coplanar. Thus, discs that do not go polar evolve as in the pure binary case. 

It is worth noting that also the binary separation plays a role in the degree of alignment of circum-binary discs (see figure \ref{fig:CzekalaPlotsma}). Indeed, short period binaries (no matter if isolated or with an external companion) present coplanar discs. The inclination distribution of that region of the parameter space could be affected also by the short period binary formation mechanisms that drive initially wider binaries to shrink. Still, wider binaries inclination distributions present the same trends discussed in the previous paragraph (i.e. highly misaligned discs around inner binaries and more coplanar discs around pure binaries).

Even though the statistics are low, in the population of discs orbiting more than two stars there are no coplanar discs, neither highly misaligned (possibly polar) ones. Indeed, the highest blue point refer to the degenerate solution with an high inclination for the GG Tau A circum-triple disc. All the hydrodynamical models of GG Tau A in the literature favour the mildly inclined solution \citep{ggtau2, ggtau3, Cazzoletti+17}. Thus, the correct tilt for GG Tau A is around 30 degrees despite the uncertainty in the astrometry and the highest blue point should be discarded.
The lack of polar discs agrees with our analytical findings. The lack of coplanar discs constrasts with our results in discs orbiting coplanar hierarchical systems. Indeed, we found that circum-triple discs evolve as orbiting a circular binary (due to the fast precession of the eccentricity vector). In addition, gived that they form via the same mechanisms, we expect the same initial tilt distribution for discs in pure binaries and in hierarchical systems. If the initial condition and the evolution are similar, then \emph{why is the tilt distribution of pure binaries and systems with more than two stars so different?}

We suggest the answer lies in the orbital dynamics of systems with more than two stars. This kind of systems have access to a richer dynamical evolution compared to pure binary systems. As previously discussed their orbital parameters and orbital orientation evolve with time on a shorter timescale compared to the disc lifetime. In particular, misaligned hierarchical systems vary their inclinations with time. Specifically, let us consider a hierarchical triple system orbited by a low-misaligned accretion disc. Looking at the pure binary population, on the long run the disc should become coplanar with the triple orbital plane. If the inner and outer orbit of the triple are misaligned, however, we expect their mutual inclination to evolve with time due to Kozai-Lidov oscillatinos. Thus, the inclination of the outer orbit will oscillate as well. As a consequence, the mutual inclination between the circum-triple disc and the triple outer orbit will oscillate with time, even if the disc had enough time to align to the stellar plane.
Thus, the observed misalignment in the systems with more than two stars population is possibly the result of these stellar orbit evolution processes, and not the outcome of accretion disc evolution. This implies that the misaligned configurations we observe in systems with more than two stars are not stable --- or slowly evolving --- configurations. Indeed, because we are taking a snapshot of an oscillating stellar orbital plane, we happen to be observing a tilted disc by pure chance.

\section{Conclusions}
\label{sec:concl}

In this work we showed that the requirement on the parameters of a circum-binary disc for going polar are necessary but not sufficient when dealing with hierarchical systems. A crucial additional requirement is that the eccentricity vector precession timescale of the system orbited by the accretion disc has to be longer than the disk libration timescale. We derived an analytical criterion to be satisfied in order for a disc to go polar around the outer levels of a hierarchical system (Eq. (\ref{eq:T2})) and around the innermost hierarchical level (Eq. (\ref{eq:T1})).

We found that discs orbiting the outer level of a hierarchical system can hardly polarly align. Except for radially narrow discs and very small mass and semi-major axis ratios, the precession of the outer orbit eccentricity vector is always faster than the nodal libration. Conversely, discs orbiting the innermost level of a hierarchical system are able to go polar as the precession of the inner eccentricity vector is slower than polar alignment. Smoothed particle hydrodynamics simulations confirm these results and surveys of circumbinary accretion discs are also in agreement --- even though statistics are still poor.

In addition, we found that the inclination of an accretion disc orbiting a hierarchical system evolves as if it was orbiting a circular binary. The disc sees the central system eccentricity vector averaged over the precession period. This is in contrast with the fact that pure binary systems and systems with more than two stars host different disc populations. The former presents discs mostly coplanar with the stellar orbital plane with little spread in the tilt distribution. The latter consists of misaligned discs with a wider spread in tilt distribution. We suggest that the scatter in misalignment observed in system with more than two stars is due to the secular oscillation of hierarchical systems orbital parameters, that continuously vary the stellar orbital plane orientation, rather than to the evolution of the accretion discs.

In conclusion, this work shows that when analysing the population of discs around binaries it is important to separate pure binaries from binaries inside hierarchical systems. Indeed, the misalignment distributions of pure binaries and systems with more than two stars tell us different stories. The former is the result of viscous disc evolution, while the second one is related to N-body orbital parameter oscillations. This additional complexity can be a precious tool to deeper investigate the impact of multiplicity on accretion disc evolution and planet formation, and, in general, to better understand the formation of multiple stellar systems. The increasing statistics of this population will allow us also to better measure the fraction of polar discs and the distribution of mutual inclination in the two populations. This is an important test for our theory. Indeed, the final distribution have to be the result of the initial conditions we derived with models, plus the theoretical expected evolution. A mismatch between the predicted and the observed populations would highlight missing pieces in the theory of disc formation and evolution.

\section*{Acknowledgements}

The authors thank the referee for their constructive feedback and suggestions, which have significantly improved the original manuscript. The authors thank KAVLI Institute of Theoretical Physics in Santa Barbara for the hospitality during the completition of this work. This project and the authors have received funding from the European Union’s Horizon 2020 research and innovation programme under the Marie Skłodowska-Curie grant agreements N. 823823 (DUSTBUSTERS RISE project), 896319 (SANDS), and 101042275 (project Stellar-MADE). CL acknowledges support from Fulbright Commission through VRS scholarship. This research was funded, in part, by ANR (Agence Nationale de la Recherche) of France under contract number ANR-22-ERCS-0002-01. SC and CL thank Daniel Elsender, Daniel Price, Matthew Bate and Yan-Fei Jiang for useful discussions. We used the following Python tools and packages: {\sc NumPy} \citep{Numpy20}, {\sc Matplotlib} \citep{Matplotlib07} and {\sc Jupyter} \citep{Jupyter16}.

\section*{Data Availability}

The data underlying this article will be shared on reasonable request to the corresponding author. The code \textsc{Phantom} used in this work is publicly available at \url{https://github.com/danieljprice/phantom}.



\bibliographystyle{mnras}
\bibliography{example} 








\bsp	
\label{lastpage}
\end{document}